\documentclass[preprint,3p,times,12pt]{elsarticle}
\usepackage{lmodern}
\usepackage{hyperref}

\usepackage{amssymb}
\usepackage{amsmath}
\usepackage{setspace}
\journal{Elsevier}

\begin{document}

\begin{frontmatter}


    \title{Computation Offloading Strategies in Integrated Terrestrial and Non-Terrestrial Networks}


    \author[nust]{Muhammad Ahmed Mohsin}
    \ead{mmohsin.bee20seecs@seecs.edu.pk}
    \author[nust]{Muhammad Umer}
    \ead{mumer.bee20seecs@seecs.edu.pk}
    \author[nust]{Amara Umar}
    \ead{aumar.dphd19seecs@seecs.edu.pk}
    \author[calgary]{Hatem Abou-Zeid}
    \ead{hatem.abouzeid@ucalgary.ca}
    \author[nust]{Syed Ali Hassan}
    \ead{ali.hassan@seecs.edu.pk}

    \affiliation[nust]{organization={School of Electrical Engineering and Computer Science (SEECS),\\National University of Sciences and Technology (NUST)}, country={Pakistan}}
    \affiliation[calgary]{organization={Department of Electrical and Software Engineering, University of Calgary}, country={Canada}}

    \begin{abstract}
        The rapid growth of computation-intensive applications like augmented reality, autonomous driving, remote healthcare, and smart cities has exposed the limitations of traditional terrestrial networks, particularly in terms of inadequate coverage, limited capacity, and high latency in remote areas. This chapter explores how integrated terrestrial and non-terrestrial networks (IT-NTNs) can address these challenges and enable efficient computation offloading. We examine mobile edge computing (MEC) and its evolution toward multiple-access edge computing, highlighting the critical role computation offloading plays for resource-constrained devices. We then discuss the architecture of IT-NTNs, focusing on how terrestrial base stations, unmanned aerial vehicles (UAVs), high-altitude platforms (HAPs), and LEO satellites work together to deliver ubiquitous connectivity. Furthermore, we analyze various computation offloading strategies, including edge, cloud, and hybrid offloading, outlining their strengths and weaknesses. Key enabling technologies such as NOMA, mmWave/THz communication, and reconfigurable intelligent surfaces (RIS) are also explored as essential components of existing algorithms for resource allocation, task offloading decisions, and mobility management. Finally, we conclude by highlighting the transformative impact of computation offloading in IT-NTNs across diverse application areas and discuss key challenges and future research directions, emphasizing the potential of these networks to revolutionize communication and computation paradigms.
    \end{abstract}

    \begin{keyword}
        Non-terrestrial networks \sep mobile edge computing \sep computation offloading \sep resource allocation \sep integrated networks \sep 6G communications
    \end{keyword}

\end{frontmatter}


\section{Introduction}

Technological advancements have propelled us into an era of unprecedented connectivity and data-driven applications. From immersive augmented reality (AR) and virtual reality (VR) experiences to autonomous driving and the widespread reach of Internet of Things (IoT), we are witnessing a significant shift in how we interact with the world. These advancements, however, bring a growing demand for computational power that pushes the boundaries of traditional computing paradigms.

Applications like AR/VR demand high-resolution graphics and real-time processing for seamless user experiences. Autonomous vehicles rely on continuous data analysis for navigation and safety. IoT connects billions of devices globally, generating vast quantities of data that require efficient processing and management to derive meaningful insights for applications such as smart cities, healthcare, and industrial automation~\cite{8552403, 9417127, 9424177, 9793590, 10068215}. Such computationally intensive applications necessitate robust and responsive networks that can handle large data volumes with minimal latency.

Existing terrestrial networks often struggle to meet these increasing demands. Their reliance on ground-based infrastructure, like cellular networks and fiber optics, inherently limits their capabilities, resulting in coverage gaps, capacity limits in densely populated zones, and insufficient latency for real-time applications. These limitations are particularly pronounced for applications that require high bandwidth and ultra-low latency, especially those driving the evolution towards 6G~\cite{9275613, 9889300, 10422712, jamshed2025tutorial, 10847914, jamshed2024non, mohsin2025deepreinforcementlearningoptimized}.

Non-terrestrial networks (NTNs) have emerged as a promising solution to complement terrestrial networks. Aerial platforms, such as high-altitude platforms (HAPs) and unmanned aerial vehicles (UAVs), along with space-based infrastructure like low-earth orbit (LEO) satellites, can overcome geographic limitations and serve remote regions~\cite{9889300, nauman2023100986, zhang2023satellite, mohsin2025hierarchicaldeepreinforcementlearning}. Integrating these platforms is widely recognized as a key enabler for 6G networks, with the potential to reshape communication paradigms~\cite{9275613, 10494748}. The integration of aerial platforms for computation offloading has also become a crucial strategy to address the resource limitations of mobile devices and utilize distributed computing capabilities. Offloading computationally intensive tasks from resource-constrained devices to dedicated remote servers or edge nodes located on these platforms can significantly reduce latency, enhance energy efficiency, and improve the overall quality of service (QoS)~\cite{10382630, 10273419, 9417127}.

This chapter explores various computation offloading strategies within integrated terrestrial and non-terrestrial networks (IT-NTNs). We provide a comprehensive analysis of how these networks manage computational workloads, allocate resources, and ensure efficient and reliable service delivery. We explore the key technologies and algorithms that underpin these strategies and discuss their strengths and limitations. We also examine potential use cases of computation offloading across diverse applications and present future research opportunities. By examining the complexities and possibilities of this technology, we aim to provide a roadmap for realizing the full potential of IT-NTNs in the next generation of communication networks.

\section{Background}

The convergence of mobile edge computing (MEC) and IT-NTNs is reshaping computation and communication, particularly for devices with limited resources and in challenging environments. This section offers a detailed look at MEC, its development, and its integration within IT-NTNs.

\subsection{Mobile Edge Computing}

MEC is revolutionizing modern networking by bringing data storage and computation closer to user equipment. This significantly reduces latency for applications that demand quick response times, such as AR, VR, autonomous driving, and the industrial IoT (IIoT). By strategically positioning data centers near users, MEC avoids the delays inherent in long-distance data transmission to centralized cloud servers~\cite{hartmann2022edge,sodhro2019308}.

The evolution of MEC towards multiple-access edge computing has further expanded its capabilities. Supporting various access technologies, including Wi-Fi, 5G, and satellite networks, has enhanced MEC's versatility and adaptability~\cite{9240934}. This allows MEC to cater to diverse application needs across different network environments. Multiple-access edge computing also offers greater flexibility in resource allocation and enables dynamic adjustments based on real-time network conditions, thereby enhancing network performance, reliability, and supporting a wider range of applications, from high-resolution streaming to complex data analytics.

MEC offers significant advantages for resource-constrained devices (RCDs) through computation offloading. This involves transferring the processing tasks from RCDs, like IoT sensors and wearable electronics, to more powerful edge servers. Computation offloading reduces latency and enhances energy efficiency at the device level. Efficient computation offloading process requires careful consideration of multiple factors like task complexity, latency requirements, energy constraints, available computing resources, and associated
costs.

\subsection{IT-NTNs for Computation Offloading}

IT-NTNs for computation offloading offer a powerful approach to overcome the limitations of purely terrestrial MEC deployments. This integration allows for the seamless convergence of terrestrial networks with NTN, capitalizing on the strengths of both to provide ubiquitous connectivity, enhanced capacity, and reduced latency for computation-intensive applications, especially in remote or challenging environments~\cite{9793590, 10382630}.

A typical IT-NTN architecture incorporates a multi-tiered structure comprising terrestrial base stations, UAVs, HAPS, and LEO satellites. Terrestrial base stations act as the foundation, offering high-speed, low latency communication in densely populated areas, while non-terrestrial platforms extend coverage to remote or underserved regions. Specifically, LEO satellites provide global connectivity and enable communication in areas where terrestrial infrastructure is limited or non-existent. HAPS are generally positioned in the stratosphere and offer wide-area coverage by acting as relay stations, thereby enhancing connectivity and offloading traffic from congested terrestrial networks. Similarly, UAVs provide on-demand coverage, proving particularly valuable in temporary or dynamic scenarios like disaster relief operations or large-scale events.

This integrated architecture facilitates streamlined data transmission and computation offloading by dynamically selecting the most appropriate link for each communication session. IT-NTNs offer a versatile and resilient infrastructure that can adapt to changing network conditions, user mobility, and application requirements. Utilizing the distributed nature of these platforms, IT-NTNs can enable efficient offloading of computationally intensive tasks and data processing, improving the overall performance and reliability of the network.

In the following sections, we expand on the computation offloading strategies within IT-NTNs and explore the enabling technologies and algorithms that drive these strategies.

\section{Edge Computing Strategies in IT-NTNs}

Optimizing performance for resource-constrained devices is a crucial aspect of modern networking. Computation offloading offers a powerful solution by shifting computationally demanding tasks from these devices to more capable servers or networks. This section explores different computation offloading strategies within IT-NTNs, highlighting their benefits, limitations, and suitability for various application scenarios.

\subsection{Edge Offloading}

Edge offloading involves shifting computational tasks from end devices to nearby edge servers strategically placed within the network. These servers might be located at roadside units (RSUs) or at other intermediate nodes close to users. The primary benefit of edge offloading is minimizing latency by reducing the distance data travels for processing, which is especially advantageous for applications demanding real-time processing. Edge offloading also reduces congestion and bandwidth usage on the core network by processing and aggregating data locally. This frees up resources for other critical services. Local data transmission enhances privacy and security by reducing the exposure of sensitive information to broader network vulnerabilities.

Despite these advantages, edge offloading has several limitations. Edge servers typically have less computational power and storage capacity than cloud servers, potentially restricting task complexity. The distributed nature of edge infrastructure also presents management and maintenance challenges. Nevertheless, edge offloading remains promising for applications requiring low latency and localized data processing, particularly in scenarios with limited or unreliable cloud connectivity.

\subsection{Cloud-based Offloading}

Cloud offloading transfers computational tasks from mobile devices to remote cloud servers via terrestrial or NTNs backhaul~\cite{7463066}. This strategy leverages the substantial processing power and storage capacity of cloud servers, delivering significant performance boosts for mobile applications while conserving the limited resources of mobile devices. Cloud offloading is well-suited for applications requiring complex computations, access to large datasets, or long-term data storage, such as machine learning model training, scientific computing, and big data analytics.

Cloud offloading also faces challenges in the form of increased latency arising due to the physical distance between devices and remote servers. Network connectivity issues can affect reliability and thus require careful planning. Security and privacy concerns related to data transmission and storage in the cloud must also be addressed.

\subsection{Hybrid Offloading}

Hybrid offloading combines the strengths of edge and cloud computing by dynamically allocating computational tasks based on their specific needs \cite{9344666, 10589437}. This approach offers flexibility by offloading tasks either to edge nodes for low latency or to more powerful cloud servers for ample resources, albeit with higher latency. Decisions are guided by factors such as latency sensitivity, computational intensity, and data privacy concerns. This enables efficient resource use, reduces latency for time-critical tasks, and enhances overall system performance. Hybrid offloading is particularly beneficial for applications with diverse requirements, such as video analytics. Initial processing can occur at the edge for real-time object detection, while more complex analysis and long-term storage can be handled in the cloud.

Implementing hybrid offloading requires effective task scheduling algorithms to distribute tasks efficiently between edge and cloud environments. Load balancing is also required to prevent node overload. As with cloud offloading, addressing security and privacy concerns arising from distributing tasks across multiple computing layers is also essential.

\subsection{Deployment Models}

Edge computing deployment models within IT-NTNs are more diverse and dynamic than traditional terrestrial systems due to the integration of non-terrestrial platforms. While concepts like network edge, regional edge, on-premise edge, and on-device edge still apply, their implementation and characteristics are highly influenced by the attributes of each NTN component. This flexibility allows edge computing deployments to be tailored to specific application needs and environmental constraints.

In network edge deployments, edge servers can be located not only at terrestrial base stations but also on UAVs, HAPs, or LEO satellites. This expanded reach of edge computing enables services in remote areas and supports applications like real-time data processing for autonomous vehicles or remote patient monitoring. Regional edge deployments can leverage local data centers interconnected with satellite gateways or terrestrial backhaul links, bringing computation closer to users in regions with limited terrestrial infrastructure. On-premise edge deployments in IT-NTNs might involve setting up edge servers at remote monitoring stations or industrial facilities, using satellite or UAV connectivity for backhaul and data transmission. Similarly, on-device edge computing in this context can utilize the processing capabilities of devices onboard UAVs or other mobile platforms, enabling localized data processing and analysis for applications like precision agriculture or environmental monitoring.

These diverse deployment models offer a wide range of offloading options within IT-NTN architectures. The most suitable model depends on the application's specific requirements, the available network infrastructure, and security considerations. For example, delay-sensitive applications may benefit from network edge or on-device computing, while computationally intensive tasks might be better suited for regional edge or cloud offloading.

Building upon these deployment models and offloading strategies, the next section explores the enabling technologies and algorithms that facilitate efficient computation offloading in IT-NTNs.

\section{Enabling Technologies and Algorithms}

Cutting-edge technologies and algorithms are paving the way for computation offloading in IT-NTNs. These advancements are crucial for tackling various challenges and capitalizing on opportunities presented by combining terrestrial and non-terrestrial resources. This section explores key enabling technologies and algorithms that make efficient computation offloading possible.

\subsection{Key Enabling Technologies}

Several technologies are instrumental in enabling efficient computation offloading in IT-NTNs and are considered key enablers of beyond 5G and 6G networks. We discuss some of the most prominent ones below.

\subsubsection{Novel Multiple Access Schemes}

Efficient spectrum use is vital for supporting massive connectivity and enabling numerous devices to offload computation tasks simultaneously. Non-orthogonal multiple access (NOMA) boosts spectral efficiency by enabling multiple users to share the same radio resources, such as time and frequency~\cite{8972353}. Unlike orthogonal multiple access (OMA), which allocates orthogonal resources to avoid inter-user interference, NOMA uses superposition coding at the transmitter and successive interference cancellation (SIC) at the receiver. By leveraging power domain differences and varying channel conditions, NOMA can accommodate more users within the same resource block, significantly improving spectral efficiency. In IT-NTNs, NOMA can enable simultaneous offloading from multiple ground users to a cluster of UAVs acting as decode-and-forward relays in a satellite-aerial-terrestrial network~\cite{9839187}. These UAVs can use coordinated multi-point (CoMP) transmission to cooperatively serve users on the same resource block, further enhancing coverage and system throughput.

Rate-splitting multiple access (RSMA) is another advanced multiple access scheme that builds on NOMA principles, offering a more generalized and flexible multiple access framework~\cite{9831440}. RSMA divides user messages into common and private parts, allowing for flexible interference management and a better balance between user fairness and overall system throughput. In downlink RSMA, all users decode the combined common parts of all user messages, while private parts are decoded individually using SIC. For uplink RSMA, each user's message is split into two sub-messages, and the receiver uses SIC to decode all sub-messages based on their respective power levels and channel conditions~\cite{9257190}.

Compared to NOMA, RSMA offers several advantages, particularly in uplink scenarios where it can achieve the full capacity region of the multiple access channel~\cite{9257190}. By tuning rate-splitting parameters and the decoding order, RSMA effectively manages interference, enhances user fairness, and improves the overall computation offloading rate in RSMA-aided MEC (RSMA-MEC) systems. Notably, RSMA generalizes NOMA in uplink transmission, providing greater design flexibility and potential for superior performance~\cite{9852986}.

\subsubsection{Millimeter Wave (mmWave) and Terahertz (THz) Communications}

mmWave and THz communication technologies operate at extremely high frequencies, offering abundant bandwidth and enabling very high data rate communication~\cite{10422712}. This vast bandwidth is particularly advantageous for computation offloading, allowing rapid and efficient transmission of large data volumes, resulting in lower transmission latency, which is much needed for delay-sensitive applications.

These technologies are especially well-suited for short-range, high-capacity links in IT-NTNs. For example, mmWave links can connect terrestrial base stations with UAVs and facilitate high-speed data transfer for computation offloading. THz offers even higher bandwidth and data rates, making it ideal for ultra-high-definition video streaming, real-time gaming, and other data-intensive applications. Thus, by leveraging these communication technologies, IT-NTNs can provide high-speed, low latency connectivity for computation offloading, significantly enhancing the performance of resource-constrained devices.

However, mmWave and THz signals are more susceptible to higher propagation losses and blockage by obstacles compared to lower frequency signals. This necessitates careful network planning and deployment, especially in IT-NTN scenarios where signal propagation can be affected by atmospheric conditions, terrain, and user mobility.

\subsubsection{Reconfigurable Intelligent Surfaces (RIS)}

RIS technology enhances wireless communication performance and coverage by intelligently controlling the propagation environment~\cite{9360709,9424177}. RISs are thin, reconfigurable surfaces composed of numerous passive reflecting elements that can be dynamically adjusted to reflect, refract, or absorb incoming signals, effectively creating a controllable wireless environment.

Strategic placement and configuration of RISs in IT-NTNs can significantly improve signal strength, mitigate interference, and enhance coverage. For instance, RIS mounted on a building can redirect signals from a HAP to a ground user in a dead zone where direct communication is blocked~\cite{9903387}. Similarly, UAV-deployed RISs, known as aerial RISs (ARIS), provide flexible and on-demand coverage enhancement for dynamic vehicular networks.

Complementing these enabling technologies are intelligent algorithms that optimize resource allocation, guide task offloading decisions, and manage mobility effectively within the IT-NTN framework.

\subsection{Offloading Algorithms}

Efficient computation offloading in IT-NTNs relies on intelligent algorithms that address the unique challenges of heterogeneous resources, dynamic environments, and diverse service requirements. These algorithms focus on optimizing resource allocation, guiding task offloading decisions, and managing mobility effectively.

\subsubsection{Resource Allocation}

Optimal resource allocation is crucial for maximizing the efficiency of computation offloading and meeting diverse QoS demands. It involves strategically distributing communication and computation resources among multiple users and tasks, considering factors such as channel conditions, task characteristics, and resource constraints. Resource allocation in IT-NTNs encompasses strategies for bandwidth allocation, power control, and server scheduling to ensure efficient resource use.

Consider a NOMA-MEC enabled aerial-terrestrial network, where multiple ground users offload computationally intensive tasks to a HAP equipped with an MEC server. This scenario utilizes both NOMA, a multiple access technique, and MEC, an edge computing paradigm, as previously discussed. To facilitate efficient offloading, users are grouped into clusters, with devices in each cluster offloading their data to the MEC server using the NOMA principle~\cite{10622807}. In this context, power and computational resources need to be allocated carefully to each user within the NOMA cluster to optimize performance.

\begin{figure}[t]
    \centering
    \includegraphics[width=0.675\textwidth]{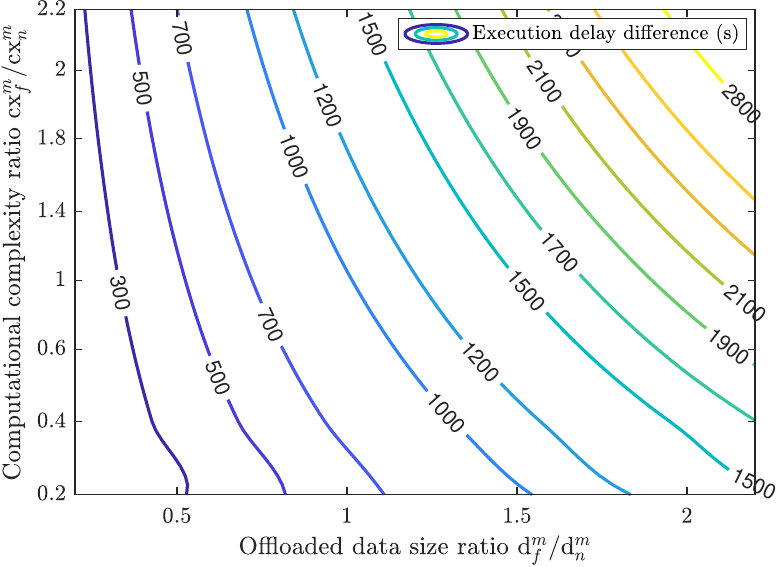}
    \caption{Execution delay difference without optimization in NOMA-MEC enabled aerial-terrestrial network~\cite{10622807}.}
    \label{fig:exec_time_no_opt}
\end{figure}

\begin{figure}[t]
    \centering
    \includegraphics[width=0.7\textwidth]{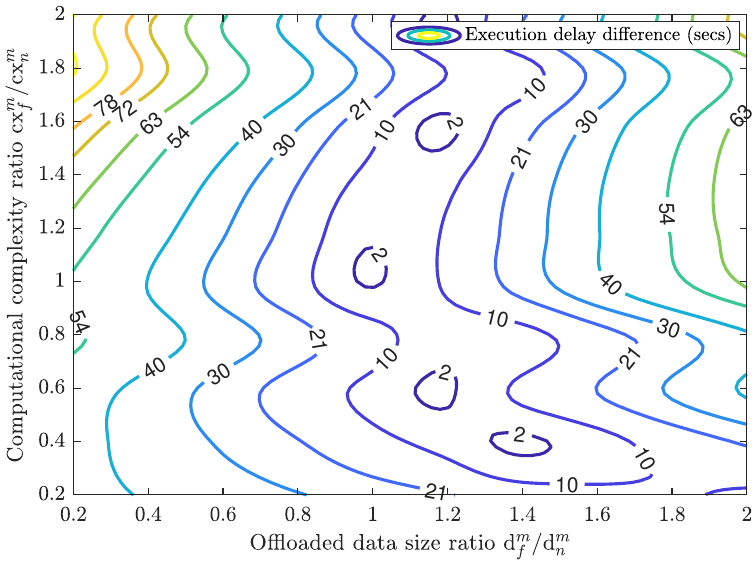}
    \caption{Execution delay difference with power and cores optimization in NOMA-MEC enabled aerial-terrestrial network~\cite{10622807}.}
    \label{fig:exec_time_opt}
\end{figure}

Minimizing the execution delay difference between the paired NOMA users is crucial for enhancing spectral efficiency and resource utilization. A significant disparity in execution delay between users leads to wasted resources, as the system remains idle while waiting for the slower user to complete its task. This inefficiency is reflected in Fig.~\ref{fig:exec_time_no_opt}, where a large difference in data offloaded by paired users, represented by the ratio $\mathrm{d}_{f}^{m} / \mathrm{d}_{n}^{m}$, results in a significant execution delay difference for different values of computation complexity ratio $\mathrm{cx}_{f}^{m} / \mathrm{cx}_{n}^{m}$.

By allocating optimal power to each user, we can equalize their data transmission times, minimizing the execution delay difference and improving overall system throughput~\cite{10622807}. Optimizing the allocation of computation resources, such as the number of cores assigned to each user at the MEC server, can also reduce the delay difference via equalization of computation times. Fig.~\ref{fig:exec_time_opt} shows that the proposed power and core optimization scheme in~\cite{10622807} effectively minimizes the execution delay difference between paired NOMA users.

To address these complex optimization problems in IT-NTNs, various techniques can be employed. For example, the Lagrange multipliers method decouples the problem into independent subproblems with their own constraints. The Karush-Kuhn-Tucker (KKT) conditions are then applied to determine the optimal values of the Lagrange multipliers, and iterative updates using a sub-gradient approach lead to a near-optimal solution. Additionally, data-aware NOMA clustering schemes, as proposed in~\cite{umar2024110335}, can further improve the effective throughput and spectral efficiency of the system.

\subsubsection{Task Offloading Decision-Making}

Another fundamental aspect of computation offloading in IT-NTNs is deciding which computation tasks to offload and to which computing resources. These decisions should be based on factors such as task complexity, latency requirements, energy constraints, available computing resources, and associated costs.

Researchers have proposed multiple methodologies to address task offloading decision-making. Optimization methods, including linear and convex optimization, model the problem by defining an objective function, like minimizing latency or energy, and corresponding constraints~\cite{10589437}. While these techniques provide optimal solutions under given conditions, they often require complete system information and can be computationally demanding for large-scale networks.

Game theory offers an alternative approach, modeling strategic interactions among multiple rational decision-makers competing for shared resources~\cite{sohrabi2020survey}. The task offloading problem can be framed as a non-cooperative game, with each user seeking to minimize their own costs, such as latency or energy~\cite{sym14030564}. Characterizing the equilibrium strategies within the game can provide valuable insights into the offloading choices made by rational users.

Machine learning algorithms, particularly reinforcement learning (RL), have emerged as effective methods for making task offloading decisions in complex and unpredictable environments~\cite{9793590, umer2024deepreinforcementlearningtrajectory}. RL algorithms learn optimal offloading policies through a trial-and-error approach, mapping system states (e.g., channel conditions, task queue lengths, resource availability) to actions (e.g., offloading decisions, resource allocation), and striving to maximize a cumulative reward (e.g., minimizing long-term cost or latency).

For instance, Li et al.~\cite{10273419} proposed an asynchronous federated deep RL (FDRL)-based computation offloading framework. The asynchronous model uploading in FDRL avoids network congestion and reduces the waiting delay for global model training. Similarly, Zhu et al.~\cite{9417127} proposed a deep RL-based task offloading (DRTO) algorithm that accelerates the learning process by adjusting the number of candidate offloading locations.

In IT-NTNs, RL algorithms can be designed to adapt to the unique challenges posed by dynamic channel conditions, varying user mobility, and unpredictable traffic patterns. For example, in a satellite-terrestrial edge computing network, an RL agent can learn to make real-time offloading decisions based on channel conditions, satellite availability, and task urgency, enabling devices to efficiently utilize both satellite edge computing (SEC) and urban terrestrial cloud (TC) for computing~\cite{9417127}.

\subsubsection{Mobility Management}

Efficient mobility management is essential for uninterrupted computation offloading in IT-NTNs, especially when users move between the coverage areas of different NTN nodes, such as terrestrial BSs, UAVs, HAPs, or satellites. Maintaining service continuity and enabling smooth handovers is necessary to ensure minimal performance degradation.

Mobile users in NOMA-MEC enabled aerial-vehicular networks operating at mmWave can experience rapid variations in channel conditions due to mobility, affecting the performance of both task offloading and resource allocation~\cite{umar2024110335}. For instance, as vehicles move, signal blockage from buildings can result in intermittent line-of-sight (LoS) and non-line-of-sight (NLoS) links with the serving HAP. This necessitates adaptive algorithms for both power control and core allocation to maintain stable offloading performance despite these fluctuations.

To address mobility management challenges, predictive handover mechanisms based on user mobility patterns and network topology information can be employed. Anticipating potential handover events beforehand can help in pre-allocating resources at the target node, resulting in lower handover latency. Efficient handover protocols are also crucial for facilitating seamless transfer of offloaded tasks and data between different NTN nodes. These protocols should be designed to minimize data loss during handover, ensure secure authentication and authorization of users at the new node, and maintain the overall QoS of offloaded tasks.

\section{Applications and Use Cases}

Computation offloading in IT-NTNs offers solutions to challenges that traditional terrestrial networks often struggle with. This section explores various applications and use cases, showcasing how these integrated networks can transform different domains.

\subsection{Autonomous Driving}

Autonomous vehicles require robust, low-latency communication networks for real-time data processing for critical functions like perception, path planning, and control. IT-NTNs address terrestrial network issues by providing ubiquitous coverage and reliable connectivity that are a must for ensuring safe and efficient autonomous driving, especially in complex urban environments.

Recent research has explored innovative approaches to support autonomous driving through integrated networks. For example, Yastrebova et al.~\cite{8970960} examined how aerial platforms like UAVs and HAPs can enhance terrestrial network infrastructure for autonomous vehicles, demonstrating that integrating these platforms can significantly expand coverage and reduce latency. Building on this, the authors in~\cite{10024371} investigated real-time HAP-assisted vehicular edge computing (VEC) for autonomous driving in remote areas. They showed that offloading computationally intensive tasks to HAPs allows autonomous vehicles to leverage superior processing capabilities while meeting strict latency requirements.

\subsection{Remote Healthcare}

Applications like telemedicine, remote diagnosis, and real-time patient monitoring demand reliable, low-latency communication networks for timely data transmission and analysis. IT-NTNs make these services possible in remote areas with limited or no terrestrial coverage, enhancing healthcare accessibility and quality for underserved regions.

The potential of edge computing in healthcare has been the focus of several recent studies. Hartmann et al.~\cite{hartmann2022edge} conducted an extensive review of edge computing in smart healthcare systems, highlighting that processing patient data locally on edge devices or nearby edge servers can significantly reduce latency, facilitating real-time monitoring and diagnosis. Complementing this, a window-based rate control algorithm was proposed in~\cite{sodhro2019308} to improve QoS in mobile edge computing-based healthcare applications like telesurgery. This algorithm dynamically adjusts the data transmission rate based on network conditions and client buffer size, reducing latency and enhancing QoS for remote medical procedures.

\subsection{Disaster Response}

In disaster scenarios, where terrestrial networks are often damaged or destroyed, the rapid deployment capabilities of IT-NTNs are invaluable. These networks provide a resilient communication infrastructure for emergency response teams, enabling them to coordinate rescue efforts, gather situational awareness, and deliver critical services to affected areas.

Researchers have proposed innovative architectures to leverage integrated networks in disaster response. Sun et al.~\cite{10382630} presented a three-layer post-disaster rescue computing architecture that leverages MEC and vehicular fog computing (VFC) in an aerial-terrestrial UAV network. This architecture supports collaborative processing of computation tasks by ground rescue vehicles and UAVs, ensuring efficient resource utilization and minimizing latency for crucial rescue operations. Further exploring the potential of HAPs, Nauman et al.~\cite{nauman2023100986} investigated high-altitude edge computing (HAEC) enabled IT-NTNs for 6G communications, highlighting the benefits of HAPs as quasi-static platforms that can guarantee higher endurance compared to low-altitude platforms, making them ideal for disaster relief and network offloading in emergency situations.

\subsection{Smart Agriculture}

IT-NTNs combined with computation offloading enhance smart agriculture practices. They provide farmers and analysts with timely and accurate data about crop health, soil conditions, and environmental factors, enabling informed decision-making, which in turn improves crop yields.

The application of UAVs and edge computing in agriculture has garnered significant attention. The authors in~\cite{rs13214387} explored in-depth the use of airborne remote sensing (RS) and edge intelligence for precision agriculture, highlighting the benefits of edge intelligence for data processing on resource-constrained UAV platforms. This reduces latency and enables real-time insights for efficient crop management. Similarly, the study in~\cite{uddin2021} proposed a cloud-edge-fog architecture for smart agriculture, where UAVs collect data from IoT sensors in the field and offload it to edge servers for processing. This architecture ensures low-latency data analysis and decision-making, enhancing crop monitoring and management.

\subsection{Industrial IoT (IIoT)}

The IIoT generates vast amounts of data from interconnected sensors and devices within industrial settings. IT-NTNs with computation offloading capabilities facilitate efficient data processing, analysis, and control, particularly in scenarios with limited terrestrial infrastructure coverage.

Recent research has explored innovative solutions for IIoT using integrated networks. Miao et al.~\cite{9849849} investigated drone swarm path planning for MEC in IIoT, proposing a multi-UAVs assisted MEC offloading algorithm to address the limitations of traditional fixed base stations in complex terrains. This leverages the flexibility and maneuverability of UAVs to provide cost-effective and efficient computation offloading services. In a related study~\cite{10068215}, a joint design of trajectory and offloading was proposed for energy-efficient UAV edge computing in the IIoT. Their work focused on minimizing energy consumption while considering the practical issue of UAV jittering, which induces uncertainties associated with flying waypoints.

These diverse applications demonstrate the potential of IT-NTNs across various sectors and highlight their ability to address challenges and enable new possibilities.

\section{Challenges and Future Directions}

IT-NTNs for computation offloading show immense promise, but several challenges must be addressed to fully realize their benefits. This field also presents exciting research opportunities that can drive innovation and development. This section outlines key challenges and explores promising future research directions.

\subsection{Key Challenges}

The heterogeneity, dynamism, and complexity of IT-NTNs pose unique challenges for efficient and reliable computation offloading. Tackling these challenges is essential to unlocking the full potential of these networks and enabling their widespread adoption across various domains.

\subsubsection{Resource Management}

IT-NTNs encompass diverse resources, including terrestrial base stations, UAVs, HAPs, and satellites, each with varying capabilities, coverage areas, and energy constraints. Efficiently managing these heterogeneous and dynamic resources is essential to maximize network performance and meet the QoS requirements of offloaded tasks. This requires intelligent algorithms for dynamic resource allocation, scheduling, and orchestration. For instance, in a scenario where UAVs assist in offloading tasks from ground vehicles to a HAP, an efficient scheme must consider the UAVs' limited battery life, the HAP's processing capacity, and the dynamic channel conditions between all components~\cite{10382630}.

\subsubsection{Mobility and Handover}

User mobility in IT-NTNs, particularly in scenarios involving aerial vehicles, can lead to frequent handovers between different NTN nodes. This dynamism challenges maintaining seamless connectivity and the smooth transfer of offloaded tasks and data during handover. Effective handover management involves designing efficient protocols and mechanisms to minimize service interruption and data loss. Predictive handover techniques can anticipate events based on user mobility patterns and network topology, allowing for proactive resource reservation at the target node~\cite{10623183}. In scenarios with fast-moving aerial vehicles, the impact of Doppler shift on communication links needs careful consideration and compensation using advanced signal processing techniques.

\subsubsection{Security and Privacy}

Offloading computation tasks to external servers raises concerns about data security and user privacy. The extended coverage and distributed nature of IT-NTNs make them susceptible to various security threats, including eavesdropping, data breaches, and malicious attacks. Ensuring secure and private computation offloading requires robust security measures such as data encryption, authentication, and access control. As Hartmann et al.~\cite{hartmann2022edge} emphasize, sophisticated privacy and data reduction methods are particularly crucial for edge computing in healthcare to protect sensitive patient data. Advanced cryptographic techniques, like homomorphic encryption, can enable computation on encrypted data without decryption, further enhancing data privacy during offloading~\cite{9830064}.

\subsubsection{Standardization and Interoperability}

Integrating diverse NTN technologies with terrestrial networks requires standardized protocols and interfaces to ensure seamless interoperability. The lack of standardized frameworks can hinder the development and deployment of IT-NTNs, limiting their scalability and efficiency. Therefore, industry and research communities must collaborate to develop and promote standardized protocols for communication, resource management, and security in IT-NTNs. These standards should address the unique characteristics of different NTN technologies while ensuring compatibility and seamless integration with existing terrestrial networks.

\subsection{Future Research Directions}

Integrating IT-NTNs and computation offloading opens up exciting opportunities for future research and development. Exploring these avenues can lead to significant advancements, enabling novel applications and pushing the boundaries of communication and computation capabilities.

\subsubsection{Advanced Control Algorithms}

The dynamic nature of IT-NTNs necessitates advanced algorithms that can adapt to varying channel conditions, user mobility, and task requirements in real-time. Leveraging machine learning techniques, such as reinforcement learning and deep learning, can significantly enhance the efficiency of resource allocation and task offloading decisions~\cite{9793590}.

\subsubsection{Predictive Mobility Management}

Efficient mobility management in IT-NTNs requires exploring novel architectures and protocols that seamlessly handle frequent handovers between different NTN nodes. Integrating software-defined networking (SDN) and network function virtualization (NFV) can provide flexible and dynamic control over network resources, facilitating efficient handover management~\cite{7890076}. Developing predictive handover mechanisms based on user mobility patterns and network topology can minimize handover latency and maintain service continuity.

\subsubsection{Blockchain and Distributed Ledger Technologies}

Blockchain and distributed ledger technologies (DLTs) offer promising solutions for enhancing security and privacy in computation offloading. Their decentralized, immutable nature provides a robust framework for secure data storage, authentication, and access control~\cite{9652086}. Exploring the use of blockchain and DLTs for managing trust and security in IT-NTNs can enhance data privacy and prevent malicious attacks.

\subsubsection{AI and Edge Intelligence}

Integrating AI and edge intelligence can significantly optimize various aspects of computation offloading in IT-NTNs. AI algorithms can predict future network traffic and dynamically allocate resources, minimizing latency and maximizing resource utilization~\cite{10494748}. Moreover, AI can facilitate intelligent task offloading decisions by analyzing task characteristics and resource availability, improving overall system efficiency.

\section{Conclusion}

This chapter explored computation offloading within IT-NTNs, revealing their potential to revolutionize data processing. We examined the rationale behind this integration, highlighting how non-terrestrial elements like UAVs, HAPs, and LEO satellites can overcome the limitations of traditional terrestrial networks. While challenges such as resource management, mobility, security, and standardization remain, ongoing research actively addresses these issues. We are thus moving closer to efficient computation in IT-NTNs, promising to transform various sectors through ubiquitous connectivity, enhanced capacity, and reduced latency. As our future becomes increasingly data-intensive and resource-constrained devices proliferate, computation offloading in IT-NTNs offer a compelling solution for real-time data processing and analysis. Continued development and deployment with collaborative efforts between industry and research communities will develop networks where communication and computation commonly intertwine. This integration will enable innovative applications, paving the way for a more connected and intelligent world.

\bibliographystyle{elsarticle-num}
\bibliography{references/main}





\end{document}